\documentstyle[aps,epsf]{revtex}
\begin{document}
\hyphenation{expec-ted}
\hyphenation{fer-mi-ons}

\title{Nonperturbative study of generalized ladder graphs in a
$\varphi^2\chi$ theory}

\author{Taco Nieuwenhuis\footnote{Email: \tt
T.Nieuwenhuis@fys.ruu.nl} and J.\ A.\ Tjon}

\address{
Institute for Theoretical Physics, University of Utrecht,
Princetonplein 5,\\
P.O.\ Box 80.006, 3508 TA Utrecht, the Netherlands.}

\date{Received January 31 1996}

\maketitle

\begin{abstract}
The Feynman-Schwinger representation is used to construct
scalar-scalar bound states for the set of all ladder and crossed-ladder
graphs in a $\varphi^2\chi$ theory in (3+1) dimensions. The results are
compared to those of the usual Bethe-Salpeter
equation in the ladder approximation and
of several quasi-potential equations.
Particularly for large
couplings, the ladder predictions are seen
to underestimate the binding energy significantly as
compared to the generalized ladder case, whereas
the solutions of the quasi-potential equations
provide a better correspondence.
Results for the calculated bound state wave functions are also presented.

\end{abstract}

\vspace{0.3cm}
\hspace{1.6cm}{PACS numbers: 11.10.St, 21.30.Fe }

\vspace{2.5cm}
\hspace{1.4cm}
Accepted for publication in {\sl Physical Review Letters}
\hfill THU-96/23 \hspace{1.8cm}

\twocolumn
One of the important issues in the study of a composite
hadronic system at higher energies, is the search for
practical and reliable
schemes to describe its relativistic dynamics.
Our knowledge about the relativistic two-body bound state problem in field
theory is
almost exclusively based on the application of the ladder approximation
to the Bethe-Salpeter equation (BSE) \cite{bse,nak}.
Unfortunately, the general applicability of
the ladder theory can be questioned on physical grounds.
In particular, the
so-called one-body limit does not
lead to the Klein-Gordon equation as it ought to.
Moreover, gauge invariance can not be satisfied within this approximation.
In order to recover these properties, at least
the set of all crossed ladder contributions is needed additionally
\cite{itzu,fsr,gross}.
So far, however, the study of the two-body Green function
beyond the ladder theory has been considered not feasible in
practice.
With this situation in mind, several quasi-potential equations (QPEs) have been
proposed and studied as possible candidates for an effective theory.
Both the ladder BSE as well as various QPEs have
been used in numerous studies throughout a wide range of systems, including
mesons \cite{milana,spence,htt2,gre}, small nuclei
\cite{htt1,wallace2,gross2}, few-electron atoms \cite{bse} and
positronium \cite{nak}.

In constructing the QPEs, one usually chooses the approximations
leading to them such that the above mentioned problems are,
at least partially, solved. However, due to our ignorance of the
behavior of the {\em full} BSE solutions, it is presently unclear
which of the, possibly infinite number of, QPEs provides the best
effective description.
In this connection it is clearly of interest to have
actual solutions available for cases where a larger class of graphs
than the ladder series is included in the BSE and that do not
suffer from the difficulties inherent to the
latter approximation. Such solutions may
serve as a testing ground for the various QPE descriptions.
Here we present results for the case where in addition
also the complete set of all irreducible crossed-ladder graphs
is included in the kernel
of the BSE, being the minimal set that is free from the
above problems. Self-energy and vertex corrections are not taken into
account.
The inclusion of these contributions are expected not to lead to qualitatively
different predictions \cite{taco2}.

In this letter the bound states formed by
two scalar particles $\varphi$ with mass $m$ interacting
through the exchange of a third scalar particle $\chi$ with mass $\mu$,
are determined using the Feynman-Schwinger representation (FSR)
\cite{fsr,taco2,taco,simo,dosch,dubin,feyhibbs}.
Starting from the Euclidean action for the above $\varphi^2\chi$ theory
\begin{eqnarray}
\lefteqn{
S =
\int {\rm d}^4x\left[\left(\partial_\lambda \varphi\right)^2 + m^2
\varphi^2\right.}\nonumber\\
&&\hspace{3cm}+\left.
\mbox{$\frac{1}{2}$}\left( \partial_\lambda\chi\right)^2
+\mbox{$\frac{1}{2}$}\mu^2\chi^2
+ g \varphi^2\chi\right] ,
\label{sdef}
\end{eqnarray}
we may reconstruct the bound state of two
$\varphi$-particles with the set of one-meson-exchange and
all irreducible crossed-ladder graphs as driving force, by explicitly
integrating out the fields in the two-body Green
function $G$. Details of this procedure can be found in Ref.\ \cite{fsr}.
According to \cite{fsr}, the FSR offers a closed expression
for the `quenched' $G$ (i.e., neglecting the possible occurrence of
vacuum fluctuation $\varphi\varphi$-loops) in terms of path
integrals over the particle trajectories $z$ and
$\bar{z}$ of the two $\varphi$ particles.
Neglecting also the contributions corresponding to the
self-energy and vertex corrections, it has the form
\begin{eqnarray}
\lefteqn{
G = \int_0^\infty\!\!{\rm d} s \:\int_0^\infty\!\!{\rm d} \bar{s}
\:\int ({\cal D} z)_{xy} \!\int ({\cal D} \bar{z})_{\bar{x}\bar{y}}}\nonumber\\
&&
\hspace{2.0cm}\exp\left( -K[z,s]-K[\bar{z},\bar{s}] + V[z,\bar{z},s,\bar{s}]
\right) ,
\label{gfsr}
\end{eqnarray}
where $K$ and $V$ are given by
\begin{eqnarray}
K[z,s] & = & m^2s +\frac{1}{4s}\int_0^1\!\!{\rm d}\tau\:\dot{z}_\lambda^2(\tau
) ,
\label{kdef}\\
V[z,\bar{z},s,\bar{s}] & = & g^2 s\bar{s}\int_0^1\!\!{\rm d}\tau\int_0^1
\!\!{\rm d}\bar{\tau}\;\Delta \left( z(\tau )-\bar{z}(\bar{\tau})\right) .
\label{vdef}
\end{eqnarray}
Our main objective here will be to compare the
predictions obtained from Eqs.\
(\ref{gfsr}-\ref{vdef}) to those from the ladder BSE and various
QPEs.

The functional integrations are over all possible paths, subject to the
boundary conditions $z(0)=x$, $z(1)=y$ and similarly for
$\bar{z}$. In (3+1) dimensions the free two-point function $\Delta (x)$ is
given
by
\begin{equation}
\Delta (x) = \frac{\mu}{4\pi^2 |x|}K_1 (\mu |x|) .
\label{twop}
\end{equation}
In \cite{fsr} it was shown that for unequal masses,
Eq.\ (\ref{gfsr}) satisfies the correct one-body limit. In addition,
it was proven that, combined
with Eq.\ (\ref{vdef}), it effectively sums up all ladder and, due
to the absence of any ordering in the interaction kernel, also all
crossed-ladder contributions to $G$. Each graph of
this set is UV-finite, so that no short-distance regularization is
required.

The bound state spectrum can be determined
by studying the behavior of $G$ with respect to variations of its initial
points $(x,\bar{x})$ and final points $(y,\bar{y})$. Considering, in
particular,
large timelike separations
$T=\mbox{$\frac{1}{2}$}(y_4+\bar{y}_4-x_4-\bar{x}_4)$,
we infer from the spectral decomposition
\begin{equation}
G = \sum_{n=0}^\infty c_n \exp\left(-m_nT\right)
\stackrel{T\rightarrow\infty}{\simeq}
c_0 \exp\left(-m_0T\right) ,
\label{gasymp}
\end{equation}
that, asymptotically the Green function is dominated by the ground state
contribution.

Notice that the path integrals in Eq.\ (\ref{gfsr}) are quantum mechanical
ones. This amounts to a considerable reduction in number of degrees
of freedom as compared to, for example, putting the field action (\ref{sdef})
on a discrete 4-dimensional lattice.
As a result accurate calculations can be carried out with this
approach also for very large times $T$.

Let us now briefly discuss the traditional Bethe-Salpeter approach
\cite{bse,nak}
to the two-body bound state problem. In the ladder approximation
the wave function $\Psi$ in momentum space obeys the following integral
equation
\begin{equation}
S^{-1}(q)\Psi (q) = \frac{\rm i}{(2 \pi)^4} \,
\int{\rm d}^4q' \, V(q-q') \, \Psi (q'),
\label{bse}
\end{equation}
where $q$ is the relative momentum between the two $\varphi$ particles.
After a Wick-rotation, the free two-body propagator
$S$ and the bare interaction $V$ assume the following form in the CM-frame
\begin{eqnarray}
S (q) & = & \frac{1}{\left({\bf q}^2 +
\omega^2+m^2-\frac{1}{4}s\right)^2 + s\omega^2},
\label{gfree}\\
V(q-q') & = & g^2\frac{1}{({\bf q}-{\bf
q}')^2+(\omega-\omega')^2+\mu^2},
\label{vbare}
\end{eqnarray}
with the relative momentum $q=({\bf q} , \omega)$.
In the bound state region Eq.\ (\ref{bse}) only supports solutions for values
of the
invariant energy $\sqrt{s}$, that correspond to bound states.

Since for unequal masses Eq.\ (\ref{bse}) in
the ladder approximation does not possess the correct one-body
limit,
several modifications to it have been proposed. Generally,
they reduce the description from a 4-dimensional to a 3-dimensional
one by making an ansatz for one of the functions involved.
This ansatz is chosen such that the resulting quasi-potential
equation do\`es possess the correct one-body limit. Here we
study three particular examples: the BSLT-equation \cite{bslt}, the equal-time
(ET) equation
\cite{htt2,htt1,wallace2,wallace}
and the Gross equation \cite{gross,gross2}, which have been widely used
in
the literature.

For the BSLT equation one assumes that the pole structure
of the two-body propagator can be approximated via a
dispersion relation. Similar to the BSLT case, in the ET prescription the
interaction is usually supposed to be independent of the relative time, i.e.,
also neglecting retardation effects.
An additional term is supplied in order to include some of the crossed-box
contributions.
In doing so, the correct one-body limit is obtained in this approach.
Finally, in the Gross formalism one puts
one of the two particles on its mass-shell by hand.
These procedures lead to the following forms of $S(q)$
\begin{eqnarray}
S_{\rm QPE}(q)
& \stackrel{\rm BSLT}{=} & 2\pi\:\delta (\omega) \frac{1}{\sqrt{{\bf
q}^2+m^2}} \frac{1}{{\bf q}^2+m^2-\frac{1}{4}s},
\\
& \stackrel{\rm ET}{=} & 2\pi\:\delta (\omega) \frac{1}{\sqrt{{\bf
q}^2+m^2}}
\frac{1}{{\bf q}^2+m^2-\frac{1}{4}s}\nonumber\\
&& \hspace{2.0cm}\times\left( 2 - \frac{s}{4({\bf
q}^2+m^2)}\right)\! ,\label{sqpes} \\
& \stackrel{\rm Gross}{=} & 2\pi\:\delta
\left(\omega+\mbox{$\frac{1}{2}$}\sqrt{s}-\sqrt{{\bf
q}^2+m^2}\right) \nonumber\\
&& \times\frac{1}{4\sqrt{s}\sqrt{{\bf q}^2+m^2}}
\frac{1}{\sqrt{{\bf q}^2+m^2}-\frac{1}{2}\sqrt{s}}.
\end{eqnarray}
For all cases the delta-function allows for the elimination of the relative
energy variable $\omega$ from the description.
The ladder BSE and 3-dimensional QPEs were solved by performing a standard
partial wave decomposition, thereby factorizing the angular
variables.

The FSR solutions were obtained by discretizing the
functional integrals, according to
\begin{equation}
({\cal D} z)_{xy} \longrightarrow \left(\frac{N}{4\pi s}\right)^{2N}
\prod_{i=1}^{N-1} \int{\rm d}^4 z_i .
\label{measure}
\end{equation}
The normalization in Eq.\ (\ref{measure}) was chosen such that, when
expanded in the coupling $g^2$, the
Green function correctly reproduces the Feynman perturbation series.
In terms of the discretized variables the functionals $K$ and $V$ assume the
following form
\begin{eqnarray}
K[z,s] & \longrightarrow & m^2s+\frac{N}{4s}\sum_{i=1}^N (z_i-z_{i-1})^2 ,
\label{kdiscr}\\
V[z,\bar{z},s,\bar{s}] & \longrightarrow &\nonumber\\
\lefteqn{
\hspace{-0.5cm}\frac{g^2s\bar{s}}{N^2}
\sum_{i,j=1}^N \Delta
\left(\mbox{$\frac{1}{2}$}(z_i+z_{i-1}-\bar{z}_j-\bar{z}_{j-1})\right)} .
\label{vdiscr}
\end{eqnarray}
The discretized boundary conditions are $z_0=x$, $z_N=y$ and
similarly for $\bar{z}$.

The integral over all degrees of freedom was performed with the
Metropolis Monte-Carlo algorithm.
The ground state mass can be obtained most efficiently by computing
the logarithmic derivative of $G$ instead of $G$ itself
\begin{equation}
L(T) \equiv - \frac{{\rm d}}{{\rm d} T} \: {\rm ln}\left[ G(T)\right]
\stackrel{T\rightarrow\infty}{\longrightarrow} m_0.
\label{lt}
\end{equation}
Introducing the shorthand notation $Z$ for the full set of degrees of freedom
and putting $S[Z]\equiv K[z,s]+K[\bar{z},\bar{s}]-V[z,\bar{z},s,\bar{s}]$,
we may write $L(T)$ as
\begin{equation}
L(T) = \left.  \int {\cal D} Z \; S'[Z] \; {\rm e}^{-S[Z]}\right/\int{\cal
D}Z\; {\rm e}^{-S[Z]},
\label{aver}
\end{equation}
where the prime denotes an analytical differentiation of the
functionals with respect to the endpoint $T$.
According to Eq.\ (\ref{aver}) the ground state mass is obtained by averaging
$S'[Z]$ over an ensemble generated by the action $S[Z]$ for sufficiently
large $T$.
The FSR ground state wave function $\Psi$ can readily
be found by performing
an additional integration of $G$ in Eq.\ (\ref{gfsr}) over the spatial relative
components
${\bf r}\equiv \bar{\bf y} - {\bf y}$ of the final point and incorporating
this coordinate in the set $Z$. By keeping track of the distribution of $|{\bf
r}|$'s when computing $L(T)$, the $\bf r$-dependence
of $\Psi$ can be determined.

The convergence in $N$ was studied and the mass of
the bound state was found to become independent
of $N$ at typical values of $N=35$-$40$.
Furthermore, $mT=40$ usually sufficed for
$L(T)$ to become independent of $T$ and to reach its asymptotic
estimate (\ref{lt}). Since the integrals over $s$ and
$\bar{s}$ in Eq.\ (\ref{gfsr}) formally diverge for large values, a cutoff
$s_{\rm
max}$ had to be introduced
in order to render the functional integrals finite. No dependence on the value
of $s_{\rm max}$ was observed.

\begin{figure}
\vspace{0.2cm}
\hspace{-0.4cm}
\epsfxsize=8.2cm
\epsfysize=6.5cm
\epsffile{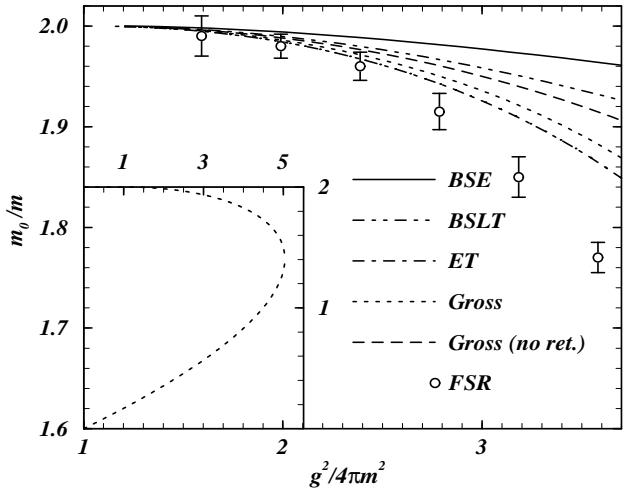}
\caption{Ground state mass $m_0$ of the $\varphi^2\chi$-theory as a
function of the dimensionless coupling constant $g^2/4\pi m^2$ for
$\mu/m=0.15$.
The inset shows the evolution of the Gross ground state
and its unphysical branch over a larger range of couplings.}
\label{fig1}
\end{figure}

In Fig.\ \ref{fig1} we present calculations of the ground state mass
as a function of the conventional (dimensionless) coupling constant $g^2/4\pi
m^2$ for the case $\mu /m = 0.15$.
\noindent
Since the self-energy contributions have been neglected in the FSR
calculations, we may directly compare the predictions to those of the ladder
BSE
and the various QPEs. The range of validity of the ladder
theory is seen to be restricted to the region of small couplings.
Generally speaking, for stronger
couplings all approximations tend to underbind the system as compared
to the FSR results. All QPEs
generate more binding energy than the ladder BSE and their results are
generally closer to the FSR ones. For the Gross equation
we also performed a calculation where the retardation in the interaction
was neglected, i.e., we simply put $\omega = \omega '= 0$ in the
potential
(\ref{vbare}). From Fig.\ \ref{fig1} we see that in this case the retardation
leads
to
additional attraction. Particularly the ET approximation is seen to
give results that relatively provide the best correspondence with the FSR ones.

We remark that due to the energy dependence in the two-body propagator,
the Gross equation allows for a second, unphysical
solution that starts at $\sqrt{s}=0$ for $g^2=0$ and for which $\sqrt{s} $
grows with increasing $g^2$. This feature is an artefact of this particular
approximation and has also been observed in
other but similar dynamical equations \cite{milana,greenberg}.
Inclusion of negative energy propagation effects was seen to cure this
pathological effect.
Both the physical and the unphysical solutions are shown
in the inset of Fig.\ \ref{fig1} and it is seen that they `annihilate'
each other at $g^2/4\pi m^2 \simeq 5.1$, for which $\sqrt{s}\simeq 1.4m$.

\begin{figure}
\vspace{0.2cm}
\hspace{-0.4cm}
\epsfxsize=8.2cm
\epsfysize=6.5cm
\epsffile{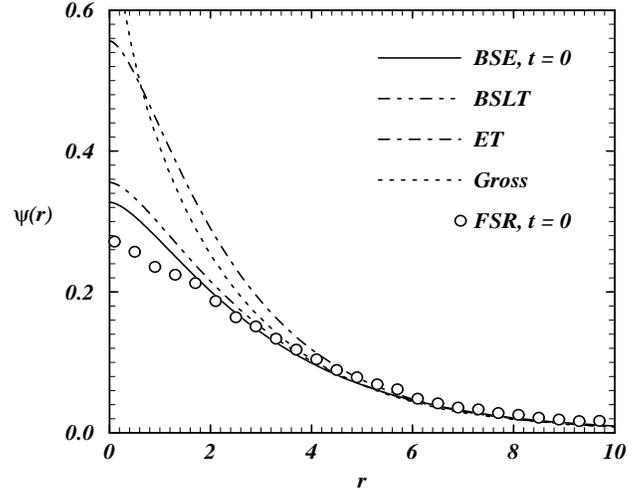}
\caption{Equal-time FSR and ladder BSE Euclidean wave function compared to
those of the
various QPEs. All solutions correspond to a bound state at $m_0 = 1.882m$.}
\label{fig2}
\end{figure}

In order to compare the FSR ground state wave function to those of the
ladder BSE and the various QPEs, we adjust the coupling constants
such that the same value of the ground state mass is found.
In Fig.\ \ref{fig2} we show the ladder BSE and FSR wave functions for relative
time $t=0$ and compare them to the QPE wave functions.
\noindent
For convenience, the FSR wave function is normalized according to
the standard nonrelativistic one.
The mass of the ground state
for all cases is $m_0 = 1.882 m$ and we take $\mu /m =0.15$.
At large separations we expect that the wave function behavior is
essentially determined by the binding energy of the composite system.
This is in agreement with the calculated results shown in Fig.\
\ref{fig2}. The main difference between the QPE predictions for short
distances is due to the asymptotic behavior of their 2-particle
free propagator $S_{\rm QPE}(q)$ for large values of $q$.

\begin{figure}
\vspace{0.2cm}
\hspace{-0.4cm}
\epsfxsize=8.2cm
\epsfysize=6.5cm
\epsffile{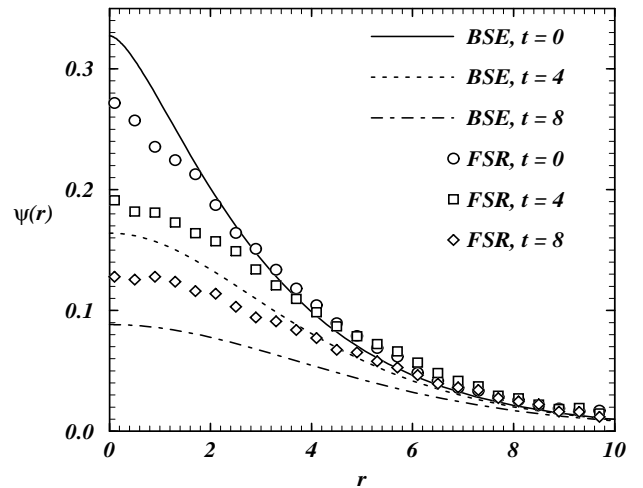}
\caption{FSR and ladder BSE Euclidean wave functions at three values of the
Euclidean relative time $t$. The ground state mass for both
calculations was at $m_0 = 1.882 m$.}
\label{fig3}
\end{figure}
Effects of relativity in the dynamics are anticipated to play an important role
in the relative time $t$ dependence of the wave function, especially at small
spatial separations between the constituents.
In Fig.\ \ref{fig3} we compare the ladder BSE and the FSR ground state wave
functions
for three values of the Euclidean relative time $t$. From this we see that
the ladder BSE prediction falls off considerably faster in
$t$ as compared to the FSR result.
This may be due to the fact that
we need a substantially larger coupling constant in the BSE case to obtain the
same binding energy. As a result the relativistic effects are enhanced
in the interaction.
At large distance both calculated wave functions agree essentially
with each other and moreover show a very slow fall off in $t$, consistent with
our expectation.

For actual hadronic systems the complication of fermions has also to be
considered.
Some progress has been achieved recently in including
spin degrees of freedom within the FSR approach. It is clearly
of great interest to study this further.
In this paper we have presented for the scalar case the first calculations of
bound state properties beyond the ladder approximation using the
Feynman-Schwinger
representation. When comparing our results to those of the Bethe-Salpeter
equation in the ladder approximation, we find that the crossed-ladders
significantly contribute to the binding energy.

\vspace{0.5cm}
\noindent
It is a pleasure to thank Yu.\ A.\ Simonov for many
valuable and illuminating
discussions concerning this work.

\end{document}